\documentclass[smallextended]{svjour3}       

\smartqed  

\usepackage{graphicx}
\usepackage{amssymb}
\usepackage{amsfonts}

\usepackage{subeqnarray}
\usepackage{amsmath}
\usepackage{subfigure}

\usepackage{multirow}
\usepackage{latexsym}
\usepackage{theorem}
\usepackage{epstopdf}
\usepackage{booktabs}
\usepackage{multirow}
\usepackage{color}
\usepackage{subfigure}
\usepackage{epsfig}
\usepackage{tikz}

\usepackage[colorlinks=false]{hyperref}

\journalname{Theoretical and Computational Fluid Dynamics}

\newcommand{\vect}  [1]{\boldsymbol{#1}}
\newcommand{\scalar}[1]{\mathsf{#1}}

\newcommand{\Ab} {\boldsymbol{A}}
\newcommand{\Bb} {\boldsymbol{B}}
\newcommand{\Cb} {\boldsymbol{C}}

\newcommand{\Tb} {\boldsymbol{T}}
\newcommand{\Kb} {\boldsymbol{K}}
\newcommand{\Lb} {\boldsymbol{L}}
\newcommand{\Xb} {\boldsymbol{X}} 
\newcommand{\Yb} {\boldsymbol{Y}} 
\newcommand{\Qb} {\boldsymbol{Q}}
\newcommand{\Rb} {\boldsymbol{R}}

\newcommand{\qv} {\vect{q}} 
\newcommand{\pv} {\vect{p}} 
\newcommand{\av} {\vect{a}} 
\newcommand{\bv} {\vect{b}} 
\newcommand{\x}  {\vect{x}}
\newcommand{\e}  {\vect{e}}

\renewcommand{\u}{\vect{\textsf{u}}}   
\newcommand{\w}  {\scalar{\textsf{d}}} 
\newcommand{\y}  {\vect{\textsf{y}}}
\newcommand{\z}  {\vect{\textsf{z}}}

\begin{document}

\titlerunning{Full-order optimal compensators for flow control.}
\title{Full-order optimal compensators for flow control: the multi-input case.}

\author{Onofrio~Semeraro \and Jan~O.~Pralits}
\institute{Onofrio Semeraro \at Department of Mechanics, Mathematics and Management (DMMM), Politecnico di Bari, Bari, Italy (\email{onofriosem@gmail.com)}
\and
Jan~O.~Pralits \at Department of Chemical, Civil and Environmental Engineering (DICCA), University of Genoa, Genoa, Italy (\email{jan.pralits@unige.it)}
}

\maketitle


\abstract{
Flow control has been the subject of numerous experimental and theoretical works. In this numerical study, we analyse full-order, optimal controllers for large dynamical systems in presence of multiple actuators and sensors. We start from the original technique proposed by \cite{2016:bewley:luchini:pralits}, the \emph{adjoint of the direct-adjoint} (ADA) algorithm. The algorithm is iterative and allows bypassing the solution of the algebraic Riccati equation associated with the optimal control problems, typically unfeasible for large systems.

\medbreak
We extend ADA into a more generalized framework that includes the design of multi-input, coupled controllers and robust controllers based on the $\mathcal{H}_{\infty}$ framework. The full-order controllers do not require any preliminary step of model reduction or low-order approximation: this feature allows to pre-assess the optimal performances of an actuated flow without relying on any estimation process or further hypothesis. 

\medbreak
We show that the algorithm outperforms analogous technique, in terms of convergence performances considering two numerical cases: a distributed system and the linearized Kuramoto-Sivashinsky equation, mimicking a full three-dimensional control setup. For the ADA algorithm we find excellent scalability with the number of inputs (actuators) in terms of convergence to the solution, making the method a viable way for full-order controller design in complex settings.
}


\newpage

\section{Introduction}\label{sec:intro}

Flow control based on linear strategies has been applied to a large variety of flows in the last decades, see \cite{1990:aber:tema}, \cite{arfm2007-kim-bewley}, \cite{fabbiane2014adaptive}, \cite{sipp2016linear}. The interest in such applications has been invigorated by the possible, numerous outcomes, ranging from drag reduction to acoustic emission mitigation. From a physical point of view, one of the main assumptions is that the modification of the coherent structures can be achieved by properly acting on the flow, both in the presence of small amplitude perturbations or large coherent structures at higher Reynolds number. If a linear approximation is valid, rigorous methods for linear control can be applied.

In this paper we focus on active control. By definition active control is characterized by energy input in the system to be controlled, by means of actuators. The system to be controlled will often be referred to as \emph{plant}. The presence of actuators allows more flexibility in the control design: sensors can be introduced such that the actuation is optimized with respect to an objective function. 

\begin{figure}
\centering
\vspace{5mm}
{\includegraphics[scale=0.35]{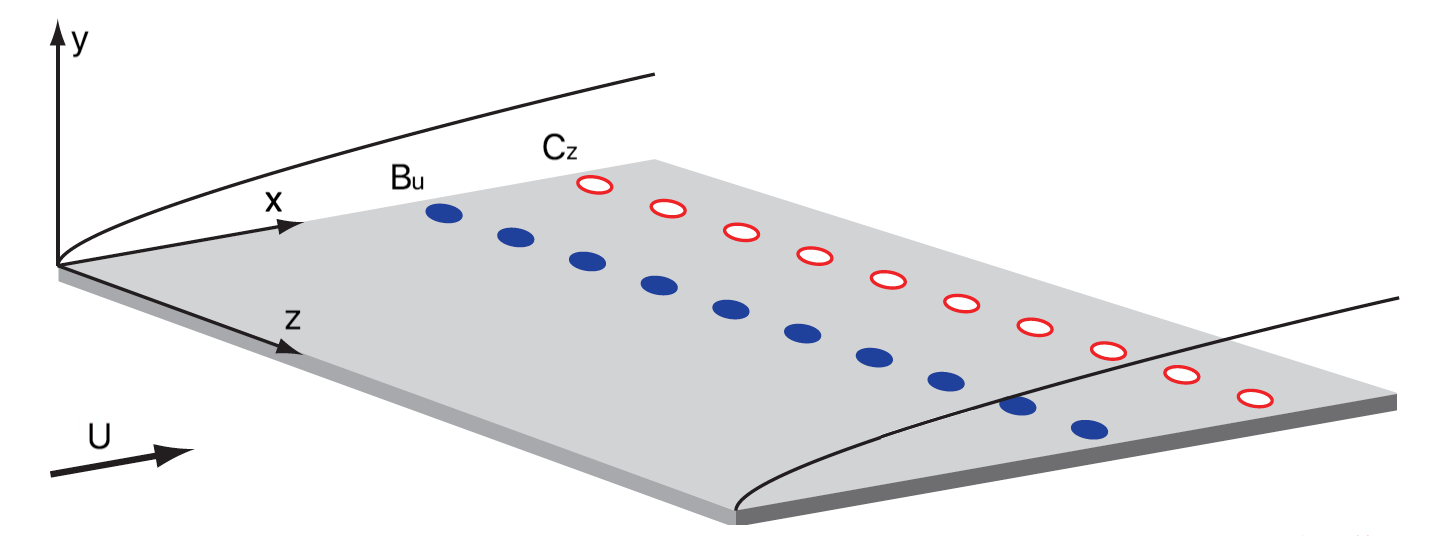}}
\vspace{5mm}
\caption{Sketch of active control for boundary layer flows; the blue dots indicate a row of localized actuators ($\Bb_{\u}$), the red circles are the sensors ($\Cb_z$) for the performance evaluation. The aim of the actuators is to modify the flow in order to fulfil a control objective. Adapted from \cite{fabbiane2014adaptive}}\label{fig1}
\end{figure}

An example of a plant to be controlled using active control strategies is given in Fig~\ref{fig1}, where the sketch of a boundary layer flow developing on a flat plate is shown, including $m$ actuators $\Bb_{\u}$ and $p$ sensors $\Cb_z$. We seek for a control law $\u(t)$, feeding the $m$ actuators. A possible way, it is to define a \emph{control kernel} $\boldsymbol{K}\in\mathbb{R}^{m\times n}$ such that the control signal $\u(t)$ -- defined in a time interval $t=[0,\, T]$ -- is proportional to the state vector $\vect{q}\in\mathbb{R}^{n}$, \emph{i.e.} $\u(t)=\boldsymbol{K}\vect{q}$. The controller is designed for fulfilling a target; in linear quadratic regulators (LQR) the following objective function to be minimized can be defined as follow
\begin{eqnarray}\label{eq:intro}
{\mathcal{J}} = \dfrac{1}{2}\int^T_{0}\left(\z^H\z +\u^H \Rb \u\right)dt.
\end{eqnarray}
The first term includes the signals recorded at the location of the sensors $\Cb_{\z}$, as $\z=\Cb_{\z}\qv$. The matrix $\Rb$ contains the control penalties for tuning the control effort (see for instance \cite{semeraro2011feedback}, \cite{semeraro2013transition} and reference therein).

A difficulty that arises in fluid mechanics is related to the dimensions of the dynamical systems: although it is relatively common to deal with numerical simulations with a number of degrees of freedom $n>10^6$, control design tools become infeasible for much smaller dimensions. A classical approach to circumvent this limitation consists in replacing the full-order system with reduced-order models capturing the \emph{essential} dynamics of the system. This methodology is sometimes called \emph{reduce-then-design}.

\subsection{Full-order control design}
Despite the necessity of reduced-order modelling for real application, the design of full-order controllers can be of great interest. Indeed, the reduced-order modelling is often performed in \emph{open-loop}, thus discarding states of the system that might be reachable in the presence of a controller; the un-modelled states can lead to inconsistencies in the control design. Moreover, when designing a controller based on reduced-order models, an estimator is often necessary: this leads to a quite larger parameters space to be explored, that includes numerous choices such as the location and the spatial distribution of the sensors/actuators pairs. In this sense, it is not always straightforward to assess whether a system is controllable and -- when possible -- if the achieved performances are the best possible due to these numerous choices that must be done during the control design. For this reason, we believe that it is of interest to pre-assess the performance of a controller by minimizing the number of these choices using optimal full-order controllers that do not require any preliminary step of model reduction or estimation processes.

Within the optimal control framework, the control gain $\Kb$ can be obtained as solution of the algebraic Riccati equation (ARE), \cite{lewis}; however, already for systems of dimensions $n\approx10^3-10^4$ the direct method is computationally intractable \cite{2008:benn:li:penz}. Neglecting the low-order design, alternative methods for the full-order design were proposed in the last decades, based on the solution of surrogate systems of equations, iterative procedures or algorithms exploiting the inherent sparsity of the considered systems. A common trait of these techniques is their feasibility when the number of actuators $m\ll n$. A classic technique is the Chandrasekhar method: the full-order ARE is replaced by a set of partial differential equations. Alternatively, the projection on low-order Krylov spaces by means of Arnoldi-type processes (see \cite{2004:benn} and references therein) and Newton methods have been proposed (\cite{1991:bank:ito}). An efficient solution method for this set of equations was proposed in \cite{akhtar2015using}, where long-time integrators are used in combination with reduced-order models based on proper orthogonal decomposition (POD) for the control of the Karman street developing behind a cylinder. Full-order kernels were computed for the control of the plane channel by \cite{hogberg:bewley:henning:03:a}, based on the works by \cite{bamieh2002}; in this method, optimal control {kernels} with a compact support in physical space are designed in the limit of parallel flow, by solving an optimal control problem for each wavenumber pair in Fourier space, independently. A Fourier anti-transform allows reconstructing the kernel in physical domain. The procedure was applied to weakly spatially developing flows by \cite{chevalier:hoepffner:akervik:henning:07} and \cite{monokrousos2008dns}, although the main drawback is the distribution of sensors/actuators pair that cannot be localized in this approach. Methods based on the pole placement are discussed, as the minimum-energy control (MCE), already introduced in \cite{2016:bewley:luchini:pralits} and \cite{bewley2007minimal}; in this limit, the control kernel is defined starting from the unstable adjoint model of the system. 
Examples are given by \cite{carini2015feedback} for the control of vortex shedding behind a cylinder, and \cite{carini2016global} for the control of the instabilities rising in the wake of a thick plate at higher Reynolds number. In the latter contribution, the linearization is performed around different mean flows, computed as RANS and U-RANS solutions.

\subsection{Iterative methods and present investigation}
Finally, iterative methods can be considered. The standard approach consists of an iteration based on the direct equation describing the system dynamics and its adjoint (see \cite{luchini2014adjoint} and citations therein). The aim of this technique is the identification of the optimal control law $\u(t)$, and it can be generalized to nonlinear settings as well as receding horizon control problem (\cite{bewley2001dns}, \cite{cherubini2013nonlinear}, \cite{kim2014adjoint}).

However, the control signal in the standard direct-adjoint iteration (hereafter indicated with DA) depends on the initial condition from which the trajectory emanates. This limitation is tackled in Bewley \emph{et al.} \cite{2016:bewley:luchini:pralits}, where the \emph{adjoint of the direct-adjoint} (ADA) algorithm is introduced, a method capable of identifying the kernel gain $\Kb$. Examples of this application can be found for the optimal control of the von K\'{a}rm\'{a}n street developing past a cylinder in \cite{pralits2010riccati} or the control of Tollmien-Schlichting waves developing in a two-dimensional boundary layer flow in \cite{semeraro2013riccati}.

\medbreak
From the theoretical point of view, the ADA algorithm replaces the original direct-adjoint optimization with the corresponding sensitivity analysis by considering the adjoint of the entire problem: this procedure changes an optimization problem of size $n$, the number of states, into a problem of size $m$, the number of inputs. The algorithm is detailed in Sec.~\ref{sec:theory}, where the original formulation is revisited. In this article we focus on two aspects of the algorithm. First, we thoroughly reconsider the design of a multi-input controller based on ADA; in Sec.~\ref{sec:multinput}, we extend the algorithm to the multi-input case, by considering the \emph{centralized} control design \cite{glad2000control}. This allows also the application of the same framework to the robust optimal control, $\mathcal{H}_{\infty}$ \cite{zhou:doyle:glover:02}. Moreover, we assess the performance of the algorithm from the convergence point of view; we compare the multi-input ADA algorithm is compared to an analogous iterative methodology introduced in \cite{maartensson2012gradient}, based on stochastic gradient (Sec.~\ref{sec:comparison}). The comparison makes use of a simple distributed system. In Sec.~\ref{sec:results}, we apply the multi-input ADA to a two-dimensional version of the Kuramoto-Sivashinsky (KS) equation. The work finalizes with conclusions in Sec.~\ref{sec:conclusions}.


\section{Linear Quadratic Regulator (LQR): iterative solution and Riccati equation}\label{sec:theory}
In this section, we concisely state the optimal control problem and derive the time-\emph{continuous algebraic Riccati equation} (CARE). For a deeper discussion, we refer to the specialized literature \cite{lewis}. We introduce the input-output equations
\begin{subeqnarray}\label{eq:odedire}
\dfrac{d \qv}{dt} &=& \Ab\qv +\Bb_{\u}\u, \qquad with \qquad \qv(0)=\qv_0,\\
\z &=& \Cb_{\z} \qv,
\end{subeqnarray}
where $\Ab\in\mathbb{Rb}^{n\times n}$ is the system matrix of dimensions $n$, the degrees of freedom. In this work, we consider time-continuous, spatially-discretized systems. The spatial distribution of the actuators is described by the matrix $\Bb_{\u}\in\mathbb{R}^{n\times m}$. The variable $\z$ is scalar and represent a time signal. If the system arises from the discretization of the Navier-Stokes equations linearized around a baseflow or a mean flow, the vector $\qv(t)\in\mathbb{R}^{n}$ represents the state of the fluid system. We want to identify a control signal $\u(t)$ such that
\begin{eqnarray}\label{eq:cost}
\mathcal{J} =
\dfrac{1}{2}\int^T_{0}\left(\qv^H \Qb \qv +\u^H \Rb \u\right)dt + \dfrac{1}{2}\qv(T)^H\Qb_T\qv(T)
\end{eqnarray}
is minimized. The matrices $\Qb\ge 0 \in\mathbb{R}^{n\times n}$, $\Rb>0 \in\mathbb{R}^{m\times m}$ and $\Qb_T\ge 0 \in\mathbb{R}^{n\times n}$ contain weights in the entries. The matrix $\Qb$ can be low-rank, for instance by defining it based on the sensors $\Cb_{\z}$ (see Fig.~\ref{fig1} and Eq.~\ref{eq:intro}). In the following, the matrices $\Qb$ and $\Rb$ are assumed diagonal, and the final condition $\Qb_T$ is set null. These assumptions do not lead to any loss of generality.

\medbreak
The solution of the control problem in Eq.~\ref{eq:odedire}--\ref{eq:cost} can be obtained by defining an augmented cost function $\mathcal{\tilde{J}}$
\begin{eqnarray}
%
%
\mathcal{\tilde J} &=& \mathcal{J} -\int^T_{0} \pv^H \left(\dfrac{d \qv }{dt} -\Ab\qv -\Bb_{\u}\u \right)dt.
\end{eqnarray}
Applying integration by parts, the following system of equations is cast
\begin{subeqnarray}\label{eq:sysopt}
\dfrac{d \qv}{dt} &=& \Ab\qv +\Bb_{\u}\u, \quad \qquad with \quad \qv(0)=\qv_0, \\
\dfrac{d \pv}{dt} &=&-\Ab^H\pv -\Qb\qv, \quad \, \, with \quad \pv(T)=0, \\
\dfrac{\partial \mathcal{\tilde{J}}}{\partial\u} &=& \Bb_{\u}^H\pv +\Rb\u.
\end{subeqnarray}
The equation for the adjoint state $\pv(t)\in\mathbb{R}^{n}$ is obtained by zeroing the gradient ${\partial \mathcal{\tilde{J}}}/{\partial\qv}$ and is integrated backward in the interval $t\in[T,0]$. The matrix $\Ab^H$ denotes the adjoint operator, satisfying the inner-product $\langle \Ab\qv,\pv\rangle = \langle \qv,\Ab^H\pv \rangle$. The unknown of the system is the control signal $\u(t)$ in the time interval $t\in[0,T]$. The solution can be approximated by iteration: at each step of the direct-adjoint iteration the control signal is updated as
\begin{eqnarray}
\u(t)^{i+1}=\u(t)^{i} -\beta^i \left( \dfrac{\partial \mathcal{\tilde{J}}}{\partial \u} \right)^i,
\end{eqnarray}
using \ref{eq:sysopt}$(c)$. The resulting control signal is optimal for a given initial condition $\qv_0$. A gradient descent algorithm can be used for defining the step $\beta$ \cite{2007:numericalrecipes}. We will refer to this technique as direct-adjoint (DA) iteration in what follows. 

\medbreak
A direct solution of the control problem is obtained by solving the associated Riccati equation. The system in Eq.~\ref{eq:sysopt} can be written in matrix form as 
\begin{displaymath}\label{eq:matrixHami}
\dot{\left(\begin{array}{c}
\qv \\ \pv \end{array}\right)}
= \left[\begin{array}{cc}
\Ab & -\Bb_{\u}\Rb^{-1}\Bb_{\u}^H \\
-\Qb & -\Ab^H\end{array}\right] 
\left(\begin{array}{c}
\qv \\ \pv \end{array}\right).
\end{displaymath} \\
Assuming the relation $\pv = \Xb \qv$, the following Riccati equation is obtained
\begin{eqnarray}\label{eq:riccati}
\Ab^H\Xb +\Xb\Ab -\Xb\Bb_{\u}\Rb^{-1}\Bb_{\u}^H\Xb + \Qb = 0,
\end{eqnarray}
for linear time-invariant system, in the steady case limit $\dot{\Xb}=0$. The matrix $\Xb\in\mathbb{R}^{n\times n}$ is the solution of the Riccati equation and it is positive-definite and symmetric. The control signal $\u(t)$ is proportional to the state $\qv$ as $\u(t) = \Kb\qv$; the constant control kernel $\Kb\in\mathbb{R}^{m\times n}$ is
\begin{eqnarray}
\Kb =-\Rb^{-1}\Bb_{\u}^H\Xb.
\end{eqnarray}
For large-system $n>10^3$ the direct solution of the Riccati equation is not feasible, due to computational costs of order $O(n^3)$ -- regardless of the structure of the system matrix $\Ab$ -- and storage requirements which are at least of order $O(n^2)$, see \cite{2004:benn}. A viable alternative is represented by iterative methods for the computation of the control kernel $\Kb$.


\subsection{Adjoint of the Direct-Adjoint (ADA) algorithm for the solution of the LQR problem}
The DA iteration does not allow the direct computation of the optimal control kernel $\Kb\in \mathbb{R}^n$; indeed, the unknowns of the problem for a given initial condition $\qv_0^i$ is the control signal $\u(t)^i$ defined in $t\in[0, T]$, with $T$ the final time of optimization. However, as observed by \cite{2016:bewley:luchini:pralits}, the following linear system can be formed
\begin{eqnarray}
\left[\u_0^1\,\, \u_0^2\,\, \dots\,\, \u_0^n \right]=\Kb_{(1\times n)},
\left[\qv_0^1\,\, \qv_0^2\,\, \dots\,\, \qv_0^n\right]_{(n\times n)},
\end{eqnarray}
where $n$ different solutions $\u(t)$ of the optimal control problem emanating from $n$ linearly independent initial conditions $\qv_0^i$ are used. Thus, $n$ iterations need to be solved, each with a different initial condition. In particular, the known vector on the left-hand side can be formed by taking the corresponding control signal at $t=0$. The columns of the matrix on the right-hand side are represented by the initial conditions of each of the $n$ iterative loops. The last step consists of the solution of a linear system of dimensions $n\times n$; so, for large system this iterative scheme is again unfeasible. 

However, as elucidated in \cite{pralits2010riccati} and \cite{2016:bewley:luchini:pralits}, one may drastically reduce the computational costs of the problem by analysing the sensitivity with respect to the initial condition using the \emph{adjoint} of the DA system. The original demonstration of the algorithm makes use of integration by-parts and it is reported in the appendix~\ref{app:A} for sake of completeness. Here, we propose an alternative version based on the properties of the Hamiltonian systems. Introducing the symplectic matrix
\begin{displaymath}
\boldsymbol{J} = \left[\begin{array}{rcccr}
\boldsymbol{0} & & &\boldsymbol{I} \\
-\boldsymbol{I} & & &\boldsymbol{0}\end{array}\right],
\end{displaymath} 
the following property is fulfilled 
\begin{eqnarray}
\Tb^H\boldsymbol{J} +\boldsymbol{J}\Tb = \boldsymbol{0}.
\end{eqnarray}
The matrix $\Tb\in\mathbb{R}^{2n\times 2n}$ for the control problem in Eq~\ref{eq:matrixHami} is defined as
\begin{displaymath}
\Tb = \left[\begin{array}{cc}
\Ab & -\Bb_{\u}\Rb^{-1}\Bb_{\u}^H \\
-\Qb & -\Ab^H\end{array}\right].
\end{displaymath}

Let us introduce the state ${\boldsymbol{x}} = \left(\begin{array}{c} {\qv} \\ {\pv} \end{array}\right)$ as solution of the system $\dot{{\boldsymbol{x}}}=\boldsymbol{T}{\boldsymbol{x}}$, with initial condition ${\boldsymbol{x}_0} = \left(\begin{array}{c} {\qv}_0 \\ {\pv}_T \end{array}\right)$ and its \emph{adjoint} state $\tilde{\boldsymbol{x}} = \left(\begin{array}{c} \tilde{\qv} \\ \tilde{\pv} \end{array}\right)$, solution of the system $\dot{\tilde{\boldsymbol{x}}}=\boldsymbol{T}^H\tilde{\boldsymbol{x}}$, with initial condition ${\tilde{\boldsymbol{x}}_0} = \left(\begin{array}{c} \tilde{\qv}_0 \\ \tilde{\pv}_T \end{array}\right)$. Note that the state ${\tilde{\boldsymbol{x}}}$ is the adjoint of the direct state ${{\boldsymbol{x}}}$ with respect to the symplectic product $\Omega(t)$, defined as
\begin{eqnarray}\label{eq:symplet}
\Omega(t) = \tilde{\x}(t)^H\boldsymbol{J}{\x}(t)=\tilde{\pv}(t)^H{\qv}(t) -\tilde{\qv}(t)^H{\pv}(t).
\end{eqnarray}
In Hamiltonian systems, this product is constant $\forall t$; including the boundary conditions, we obtain $\Omega(t)=0$. The relation \ref{eq:symplet} can be compared with the optimality condition at $t=0$
\begin{subeqnarray}\label{eq:adaopt}
\tilde{\pv}_0^H\qv_0 &=& \tilde{\qv}_0^H\pv_0, \\
\u=\Kb\qv_0 &=& (-\Rb^{-1}\Bb_{\u}^H)\pv_0.
\end{subeqnarray}
We can notice that, if we introduce as initial condition of the dual system $\tilde{\qv}_0$ one row of $-\Rb^{-1}\Bb_{\u}^H$, the resulting adjoint solution $\tilde{\pv}_0^H$ corresponds to one row of $\Kb$. Thus, the solution of the Riccati problem is obtained without solving the algebraic equation Eq.~\ref{eq:riccati}. In principle, the exact solution is obtained for $T\rightarrow\infty$. In practice, a sufficiently long time-window for the optimization guarantees convergence towards the optimal solution. Strictly speaking, the resulting problem is not anymore an optimal control problem, despite it makes use of the same equations.


\section{The multi-input version of ADA}\label{sec:multinput}
In this section we discuss how to extend the ADA algorithm to multivariable systems, \emph{i.e.} the systems characterized by the presence of multiple inputs and outputs. Within the context of flow control, this is typically the case of full, three-dimensional setups extended in the spanwise direction. An example is sketched in Fig.~\ref{fig1}. When only one actuator/sensor pair is considered, we usually refer to a Single-Input-Single-Output (SISO) system; vice versa, the opposite case is when multiple inputs/outputs are introduced, i.e. the Multi-Input-Multi-Output (MIMO) case. In the forthcoming, we consider the MIMO case. The main issues for the MIMO design are introduced in Sec.~\ref{sec:multi}. The extension of the ADA algorithm for the solution of the optimal control is discussed in section Sec.~\ref{sec:centra}; the dual iteration for the estimation is briefly introduced in Sec.~\ref{sec:esti}. Finally, the robust, optimal control design is revisited in Sec.~\ref{sec:hinfty}, where the multivariable extension of ADA is adapted for the $\mathcal{H}_{\infty}$ design (\cite{skogestad05}).


\subsection{Multivariable systems}\label{sec:multi}
From the design point-of-view, a multivariable system is potentially characterized by cross coupling between inputs and outputs. This cross-coupling is the root of difficulties in multivariable control \cite{glad2000control}: indeed, a change in one input can affect multiple outputs.

A first, simplistic approach for the MIMO design consists of designing a number of SISO closed-loop that equals the number of sensor/actuator pairs; considering the example in Fig.~\ref{fig1}, $m=p$ closed loop can be designed for each of the actuators, based on one sensor located at the same upstream location $z_0$. This approach is called \emph{decentralized} approach. In this case the number of actuators must equal the number of sensors, unless neglecting some of the elements; more importantly, the cross-couplings are disregarded. The cross-couplings may of course affect both stability and performance of the closed loop system. In general, the stability of the closed-loop is not guaranteed. However, if the decentralized controller is stable in each SISO loop and weak couplings characterize the original plant, then the closed loop is also stable. In terms of performance, the resulting controller will be sub-optimal.

The opposite approach consists of accounting all the possible couplings. This methodology is called \emph{centralized control} design. This choice guarantees optimal performances and stability of the closed loop, although it is less easy to implement in practical situations. 

\medbreak
In the following, we discuss only these two limits. In general, linking the sensor signals with the controller that have the strongest interactions allow for hybrid solutions where a number of sensors are wired with a number of actuators; this is the so-called \emph{pairing problem}.


\subsection{Centralized controllers using ADA}\label{sec:centra}
Here, we analyse the centralized and decentralized design with respect to the optimal problem. The two cases can be summarized as follows
\begin{enumerate}
\item \emph{Decentralized control}: $m$ control gains of dimension $n$ are designed independently from each other. Thus, $m$ Riccati equations are solved.
\item \emph{Centralized control}: one control gain of dimensions $m \times n$ is designed, by solving one Riccati equation.
\end{enumerate}
The application of the ADA algorithm for the decentralized case is straightforward and it is the one \emph{de facto} discussed by \cite{2016:bewley:luchini:pralits}. Each of the $m$ SISO Riccati equation is replaced by the corresponding ADA iteration, with one input and one output. As already mentioned, this case is only suboptimal and does not guarantee the stability of the closed loop.

The ADA algorithm does not allow identifying a low-rank approximation for the matrix $\mathbf{X}$; thus, the design of the centralized controller needs to be performed row-by-row, using a number of iteration-loops that equals the number of rows. The design of a control kernel of dimensions $\Kb\in\mathbb{R}^{m\times n}$ requires $m$ iteration loops. The cross-coupling are accounted for by choosing the actuators and the sensors to be coupled in the input system and optimization cost function, respectively,
\begin{eqnarray}
\dfrac{d \qv}{dt} &=& \Ab\qv +\Bb_{u,m} \u_m, \\
{\mathcal{J}} &=& \dfrac{1}{2}\int^T_{0}\left(\qv^H\Cb_{\z,p}^H\Cb_{\z,p}\qv +\u_m^H \Rb \u_m\right)dt.
\end{eqnarray}
where we indicate with $\Bb_{\u,m} $ the matrix including {all} the $m$ actuators and $\Cb_{\z,p}$ all the sensors of the original plant. The initial condition of each of the iterations is chosen taking $\tilde{\qv_0}_i^H=-\Rb_{i,i}^{-1}\Bb_i^H$, such that each of the $i$-th rows of the matrix $\Kb$ is obtained as adjoint solution of the iteration process.

This design strategy guarantees the {coupling} among the multiple inputs and corresponds to the solution of a centralized problem. In principle, following the pairing problem, one can select the pairs of sensors/actuators that share the strongest interactions \emph{a-priori}. In the limit where only one actuator is considered in each of the $m$ loops, we recover the decentralized case where each of the actuator is designed independently from each other.


\subsection{Estimation problem}\label{sec:esti}
The technique can be extended to any problem based on the solution of algebraic Riccati equations. A first example is provided by the estimation problem as shown in \cite{semeraro2013riccati} for the single-input-single-output setting. An estimator is defined as the following dynamical system
\begin{subeqnarray}\label{eq:esti}
\dot{\hat{\qv}} &=& \Ab \hat{\qv} +\boldsymbol{B}_{\u}\u -\Lb(\y-\hat{\y}), \\
\hat{\y} &=& \Cb_{\y}{\hat{\qv}}.
\end{subeqnarray}
The estimator allows to reconstruct the original state $\qv$, starting from local measurements $\y$; the state $\hat\qv$ is defined such that the error $\y-\hat{\y}$ is minimized. The estimator is driven by the error term via the matrix $\Lb$, referred to as estimation gain. The estimation gain $\Lb$ is the unknown of the associated estimation problem, obtained as solution of the following Riccati equation
\begin{subeqnarray}\label{eq:ricesti}
&\Ab\Yb +\Yb\Ab^H-\Yb\Cb_{\y}^H\boldsymbol{G}^{-1}\Cb_{\y}\Yb+\Bb_{\w}\boldsymbol{W}\Bb_{\w}^H=0, \\
&\Lb = -\Yb \Cb_{\y}^H\boldsymbol{G}^{-1}.
\end{subeqnarray}
The $\mathbf{G}$ contains in each of the diagonal entries the estimation penalty, similarly to the control case. From the mathematical point of view, a fictitious adjoint problem is cast such that the cost function
\begin{eqnarray}\label{esti_cost}
\mathcal N\left(\pv\left(\tilde{\y}\right),\tilde{\y}\right)= 
\dfrac{1}{2}\int^T_0 \left(\pv^H \boldsymbol{W} \mathbf{p} +\tilde{\y}^H \boldsymbol{G} \tilde{\y}\right)dt
\end{eqnarray}
is mimized. In this dual system, the output $\Cb_{\y}^H$ is an input of the system, while the adjoint of the inputs $\Bb^H_{\w}$ and $\Bb_{\u}^H$ play the role of the outputs \cite{bagheri2009input}. The analogy is completed by observing that the ``feedback law'' is now represented by $\tilde{\y}(t)=\Lb^H\pv(t)$. Thus, the first term of the cost function expresses the energy of the dual state $\pv$, while the second term minimizes the energy input of the feedback law.

A justification of this deterministic approach is given by \cite{arfm2007-kim-bewley}. The physical interpretation of the method is given within the stochastic framework \cite{amr2009-bagheri-et-al}. In this case, the unknown of the Riccati equation~\eqref{eq:ricesti} is the expected energy of the estimation error $\e(t)$, while $\mathbf{W}$ is the covariance of the forcing.

The optimization machinery employed for the control problem can be adopted in an analogous manner for computing the full-dimensional estimation gain $\Lb$. The full formulation is given in \cite{semeraro2013riccati}, where the algorithm is referred to as \emph{Adjoint of the Adjoint-Direct} (AAD). Also the AAD algorithm can be generalized to the centralized/decentralized case. Thus, in presence of $p$ estimation sensors $\Cb_{\y}$, $p$ estimators can be designed based only on one sensor, independently from each other, or a centralized gain can be designed by running an iteration for each of the column of $\Lb$, based on all the sensors $\Cb_{\y}$.


\subsection{Riccati solution for $\mathcal{H}_{\infty}$ problem}\label{sec:hinfty}
The optimal control framework can be extended to the solution of robust optimization problems, introducing the \emph{worst} disturbance scenario. The cost function associated with the problem is the following
\begin{eqnarray}
\mathcal{J} =
\dfrac{1}{2}\int^T_{0}\left(\qv^H \Qb \qv +\u^H \Rb \u -\gamma^{-2}\w^H\boldsymbol{W}\w\right)dt,
\end{eqnarray}
where the extra-term to maximize is the unknown, {worst} disturbance $\w$. The optimization process consists of a simultaneous optimization problem: the signal $\u$ is computed such that the a worst-case disturbance is minimized (\emph{min-max} optimization). The parameter $\gamma$ is chosen by the user. In this sense, the resulting control is sub-optimal. For values of $\gamma\rightarrow\infty$, we approach the limit of the full-order LQR problem already considered. The direct equation of the problem reads
\begin{eqnarray}
\dfrac{d \qv}{dt} &=& \Ab\qv +\Bb_{\w}\w +\Bb_{\u}\u, \qquad with \,\,\, \qv(0)=\qv_0.
\end{eqnarray}
By introducing the corresponding augmented Lagrangian, it is possible to define the adjoint equation
\begin{eqnarray}
\dfrac{d \pv}{dt} &=&-\Ab^H\pv -\Qb\qv, \qquad \qquad\,\, with \,\,\, \pv(T)=0.
\end{eqnarray}
Two optimal conditions are defined, by zeroing the respective gradients. The first condition corresponds to the optimal control signal $\u$,
\begin{equation}
\u = -\boldsymbol{R}^{-1}\boldsymbol{B}_{\u}^H\boldsymbol{p};
\end{equation}
the second condition is the worst disturbance $\w$
\begin{equation}
\w = \gamma^{2}\boldsymbol{W}^{-1}\boldsymbol{B}_{\w}^H\boldsymbol{p}.
\end{equation}
An ARE can be obtained also for this robust, optimal control problem by imposing the equivalence $\boldsymbol{p}=\boldsymbol{X}\boldsymbol{q}$, into the Hamiltonian system
\begin{displaymath}
\begin{array}{c}\boldsymbol{T}\end{array} = \left[\begin{array}{ccc}
\Ab & \quad & \gamma^2\boldsymbol{B}_{\w}\boldsymbol{W}^{-1}\boldsymbol{B}^H_{\w} -\Bb_{\u}\boldsymbol{R}^{-1}\Bb_{\u}^H
\\
-\Qb & \quad & -\Ab^H
\end{array}\right].
\end{displaymath}
The solution is obtained by solving the following problem, based on a ARE equation
\begin{subeqnarray}\label{sec:robuare}
&\boldsymbol{A}^H\boldsymbol{X} +\boldsymbol{X}\boldsymbol{A}
-\boldsymbol{X}\left(\boldsymbol{B}_{\u}\boldsymbol{R}^{-1}\boldsymbol{B}_{\u}^H
-\gamma^{2}\boldsymbol{B}_{\w}\boldsymbol{W}^{-1}\boldsymbol{B}_{\w}^H\right)\boldsymbol{X}
+\boldsymbol{Q}=0, \\
&\Kb = -\Rb^{-1}\Bb_{\u}^H \boldsymbol{X}, \\
&\boldsymbol{Y}=\gamma^{2}\boldsymbol{W}^{-1}\boldsymbol{B}_{\w}^H\boldsymbol{X}.
\end{subeqnarray}
The control gain $\Kb$ and the worst disturbance $\boldsymbol{Y}$ are computed based on the solution $\boldsymbol{X}$ of the ARE equation.

\medskip
The ADA algorithm can be applied for the full-order approximation of this problem. Considering the single-input-single-output setting, two iterations are necessary for computing, as adjoint solution at final time $T$, $\boldsymbol{K}$ and $\boldsymbol{Y}$ introducing as initial conditions
\begin{subeqnarray}
&\boldsymbol{q}_{\boldsymbol{K},0} = -\Rb^{-1}\Bb_{\u}^H, \\
&\boldsymbol{q}_{\boldsymbol{Y},0}=\gamma^{2}\boldsymbol{W}^{-1}\boldsymbol{B}_{\w}^H.
\end{subeqnarray}
Indeed, the robust problem can be seen as a generalized multi-input Riccati equation to be solved, where all the inputs are coupled together. From the algorithm point of view, a relevant difference between the standard $\mathcal{H}_2$ problem and the $\mathcal{H}_{\infty}$ is related to the optimization process: due to the simultaneous minimization/maximization underlying the $\mathcal{H}_{\infty}$, we do not seek for a minimum but for a saddle point of the objective function (see for instance \cite{gumussoy2009multiobjective}). Finally, note that the both the disturbances $\boldsymbol{B}_{\w}$ and the $\boldsymbol{B}_{\u}$ can be multiple; also in this case, a centralized approach needs to be used.


\section{Comparisons among algorithms}
This section provides numerical results for assessing the performance of the ADA algorithm in presence of multiple inputs and centralized control. We consider a toy-problem and compare ADA with a recent algorithm proposed by \cite{maartensson2012scalable}. The test-bed is mainly meant at verifying the speed of the convergence for the different algorithms.


\subsection{Full-order controllers using stochastic gradients} 
The algorithm introduced by \cite{maartensson2012scalable}, and here indicated as MR, is used for comparison. The algorithm computes the control gain by updating at each step the solution as
\begin{equation}
\Kb^{i+1}=\Kb^{i} -\beta^{i} \left(\nabla_{\boldsymbol{K}}\mathcal{\tilde{J}}\right)^{i},
\end{equation}
where $\nabla_{\boldsymbol{K}}\mathcal{\tilde{J}}$ is the gradient for the update at each step of the iteration. The basic formulation is proposed in appendix \ref{app:B}. In principle, the algorithm suffers of the same limitations of the DA iteration: the identified solution depends by the initial condition. The problem is circumvented by choosing a new initial condition randomly at {each iteration}. The methods can be interpreted as a stochastic gradient descent method; at each step of the iteration, an initial state is chosen for which the gradient is computed. Using different initial states at every iteration, the final feedback matrix will be not depend from any of the initial conditions.
\begin{table}
\begin{center}
\begin{tabular}{ c | c | c | c | p{35mm} }
\hline
Algorithm & IC & Grad. & Loops & Notes \\
\hline
\multirow{2}{*}{DA} &
\multirow{2}{*}{Random at each loop } &
\multirow{2}{*}{$m\times n_t$} &
\multirow{2}{*}{$n$} & 
{$\Kb$ is solution of a linear system.} \\
\multirow{2}{*}{ADA} &
\multirow{2}{*}{$-\mathbf{R}^{-1}_{i,i}\Bb_i$} &
\multirow{2}{*}{$m\times n_t$} &
\multirow{2}{*}{$m$} &
{Each row of $\Kb$ is solution of one iteration-loop.} \\
\multirow{2}{*}{MR} &
\multirow{2}{*}{Random at each iter.} &
\multirow{2}{*}{$m\times n$} &
\multirow{2}{*}{$1$} &
$\Kb$ is solution of one iteration-loop, $\forall m$. \\
\hline
\end{tabular}
\end{center}
\caption{The table summarizes the main features of the algorithms compared in Sec.~\ref{sec:comparison}: the column \emph{IC} indicates the initial conditions used; the third column \emph{Grad.} indicates the dimensions of the gradients used during the process of update; in the column \emph{Loops}, the number of iteration-loops required for computing the control gains of dimension $n\times m$ is indicated. In the last column, the characterizing features of the methods are highlighted.}\label{tabula}
\end{table}


\subsection{Implementation} 
The implemented routines are in prototypical form. The main idea is to analyze the scaling of the computational costs as a function of the number of actuators $m$ and compare the resulting trends. Following this rational, a simple steepest descent algorithm, where the step along the gradient direction is computed as 
\begin{equation}
\beta=\dfrac{\left(d\mathcal{J}^{i}\right)^T\,d\mathcal{J}^{i}}{\left(d\mathcal{J}^{i}\right)^T \,d\mathcal{J}^{i+1}}.
\end{equation}
The variable $d\mathcal{J}^{i+1}$ indicates the gradient at the current step of the iteration. In principle, acceleration techniques might be applied (for instance, implicit restarting or Nesterov's accelerated methods) for further improving the performance of the algorithms. Two stop-criteria are chosen: the difference of the control cost between two iterations and the norm of the gradient. The iterations stop when one of the two criteria is fulfilled below a chosen tolerance. 

In table \ref{tabula}, we highlight the main features of the algorithms. For all the cases the centralized/decentralized synthesis needs to be imposed when choosing the actuators of each single-input sub-system. In this section we only consider the centralized version of the algorithm that is the most expensive from the computational point of view. Note that for the MR algorithm the basic form is analysed, so without considering a pre-determined structure for the kernel; due to the random selection of the initial condition, the assessment of the performance is done by considering $10$ simulations for each of the analysed cases.


\subsection{Numerical example: a distributed system}\label{sec:comparison}
A toy-problem mimicking a distributed system of dimension $n=50$ is considered. The system matrix $\Ab\in\mathbb{R}^{n\times n}$ is in tri-diagonal form
\begin{displaymath}
\begin{array}{c}\mathbf{A}\end{array}
= 
\left[
\begin{array}{rrr rrr rrr}
1 &-4 & 0 & & & & & & \\
&-1 &-4 & 1 & & & & & \\
& & 1 &-4 &-2 & & & & \\
& & &\ddots&\ddots &\ddots & & & \\
& & & & 1 & -4 & -2 & & \\
& & & & & 8 & -8 & -1 & \\
& & & & & & 0 & -2 & 1 
\end{array}
\right].
\end{displaymath}
The matrix $\Bb\in\mathbb{R}^{n\times m}$ contains non-null, unitary entries only along the main diagonal. The chosen system resembles the one used in \cite{maartensson2012scalable}. The number of actuators is indicated by $m$; 10 cases are analysed, where the number of actuators is chosen as $m=[5,10,15,\dots,50]$. For the last case, $n=m$. The system is marched in time until the final time $T=20$, with $n_t=501$. The value of the final time was deemed sufficient for properly approximating the solution.

The solution of the optimal problem is computed for three different penalty kernels, namely $l=[25,50,100]$ and benchmarked against the Riccati solution, obtained in Matlab using $\texttt{care.m}$.

\begin{figure}
\centering
{\includegraphics[scale=0.4666]{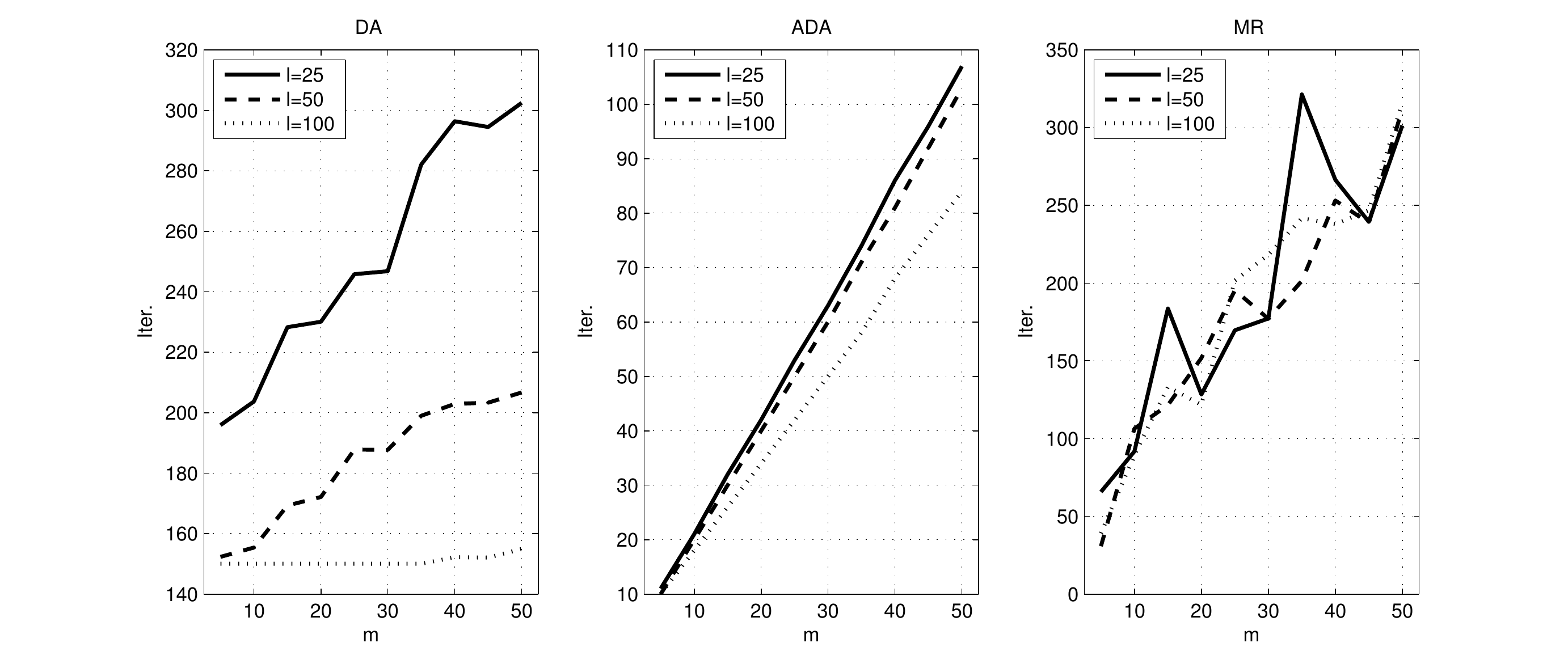}}
\caption{Convergence of the iterative algorithms. The number of iterations as a function of the number of actuators $m$ and control penalty $l$ is considered for the algorithms described in Sec.~\ref{sec:multi} and Sec.~\ref{sec:comparison}.}\label{fig2}
\end{figure}

\medbreak
In Fig.~\ref{fig2}, the number of iterations necessary for convergence is shown as a function of the number of actuators $m$. The chosen tolerance, given as the difference between the cost at $i$ and $i-1$, is $\varepsilon=1.0\times 10^{-6}$. All the algorithms scale with the number of actuators $m$. In particular, it is possible to observe a remarkable regularity for the ADA algorithm, that scales linearly with $m$; different control penalties require different number of iteration. In particular, decreasing $l$ requires a greater number of iterations. The behaviour was already observed in ADA, see \cite{semeraro2013riccati}. This same behaviour is observed in the DA algorithm; in fact, only at $l=100$ is observed a constant number of iterations with $m$, while a larger number of iterations is required with the increasing number of actuators for the solution of the problem at lower $l$.

%

The DA algorithm requires on average more iterations when compared to the other iterative methods. This is not surprising, as $n$ loops are required. However, the main bottleneck of DA is the final solution of a linear system, together with the storage of a large matrix containing along the columns the $n$ initial conditions; in that sense, the use of the DA algorithm is {unfeasible} for larger system and is considered here only for the sake of completeness of our discussion. 
%
\begin{figure}
\centering
{\includegraphics[scale=0.4666]{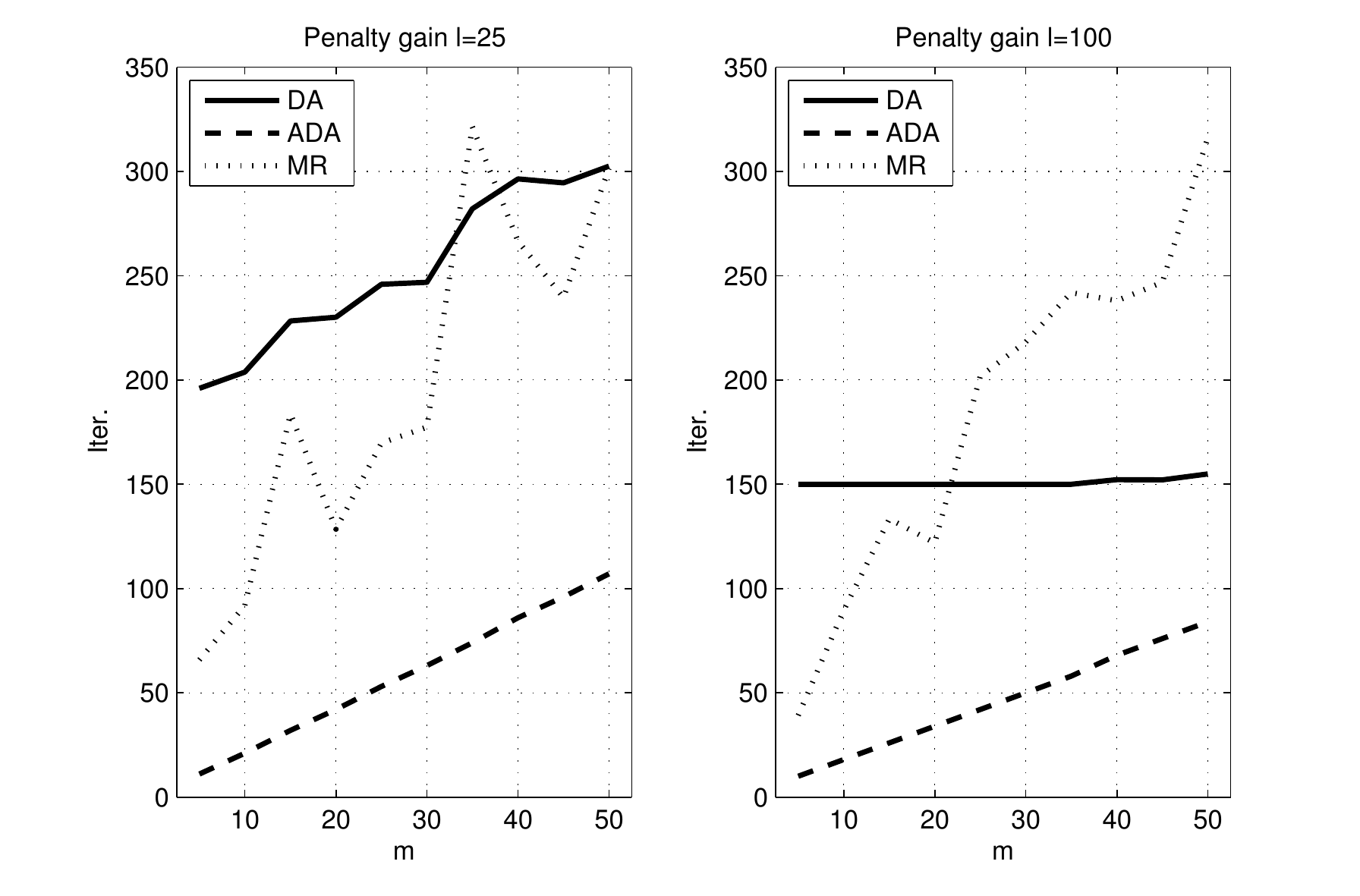}}
\caption{Convergence of the iterative algorithms. The data shown in Fig.~\ref{fig2} are considered for $l=25$ and $l=100$. The ADA algorithm is characterized by high scalability with the number of actuators and a number of iteration smaller with respect to the alternative choices.}\label{fig3}
\end{figure}

The MR algorithm has a less clear behaviour, due to the choice of the initial conditions; an increase of the number of iterations is observed with $m$ also for this case. The choice of the penalty does not affect the number of iterations required. This is clearer when observing Fig.~\ref{fig3}, where the iteration number as a function of $m$ is shown for two different penalty gains. 

In conclusion, we can summarize the performance of three algorithms as follows
\begin{itemize}
\item For all the cases, an increase of the total number of iterations with $m$ is observed. This is expected in ADA, where $m$ iteration-loop are required, while it is less obvious for the MR case, where only one iteration-loop is required regardless of $m$.
\item ADA scales linearly with $m$. This is not the case for the MR algorithm, due to the stochastic gradient application.
\item For the MR algorithm the averaged number of iterations is not affected by the control effort applied (i.e., the choice of the control penalty $l$).
\end{itemize}


\section{Numerical example: the two-dimensional Kuramoto-Sivashinsky}\label{sec:results}
\begin{figure}
\subfigure[$t=200$]{\includegraphics[scale=0.45]{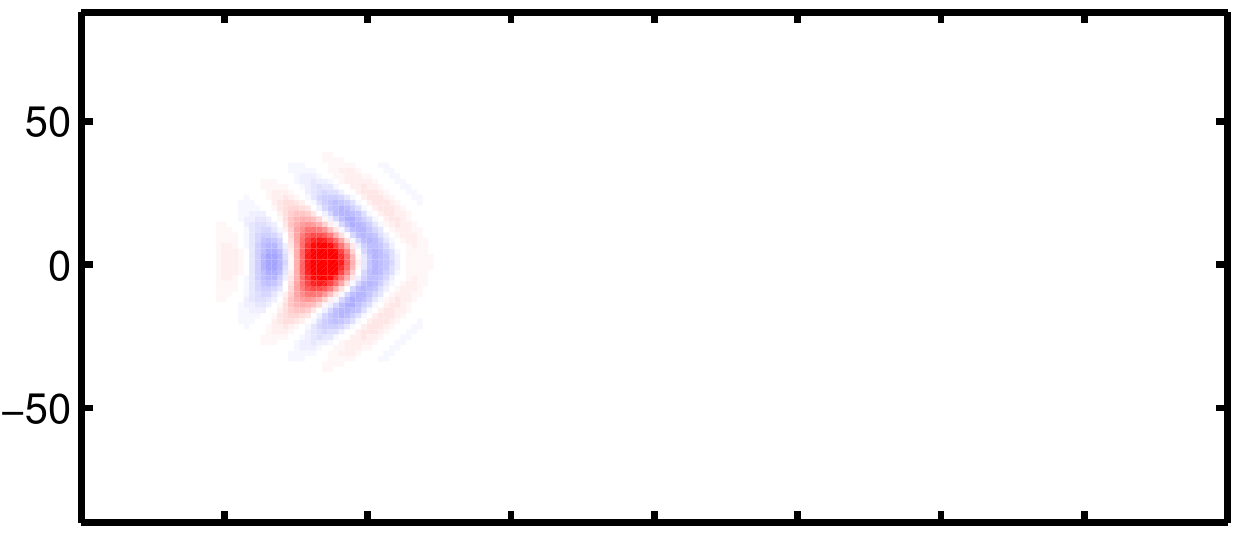}}
\hspace{.25cm}
\subfigure[$t=200$]{\includegraphics[scale=0.45]{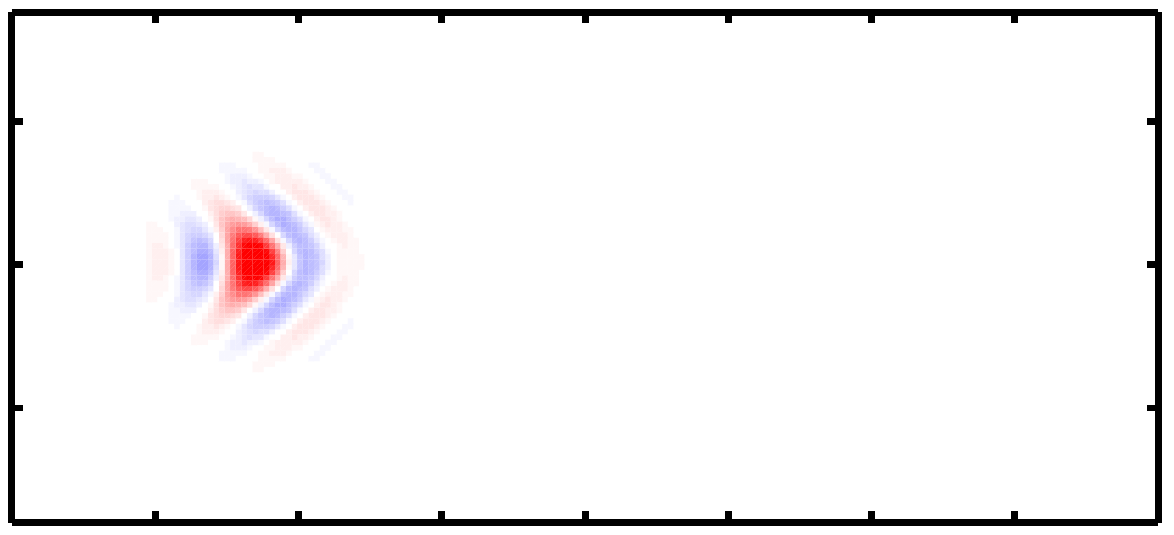}}
\\
\subfigure[$t=500$]{\includegraphics[scale=0.45]{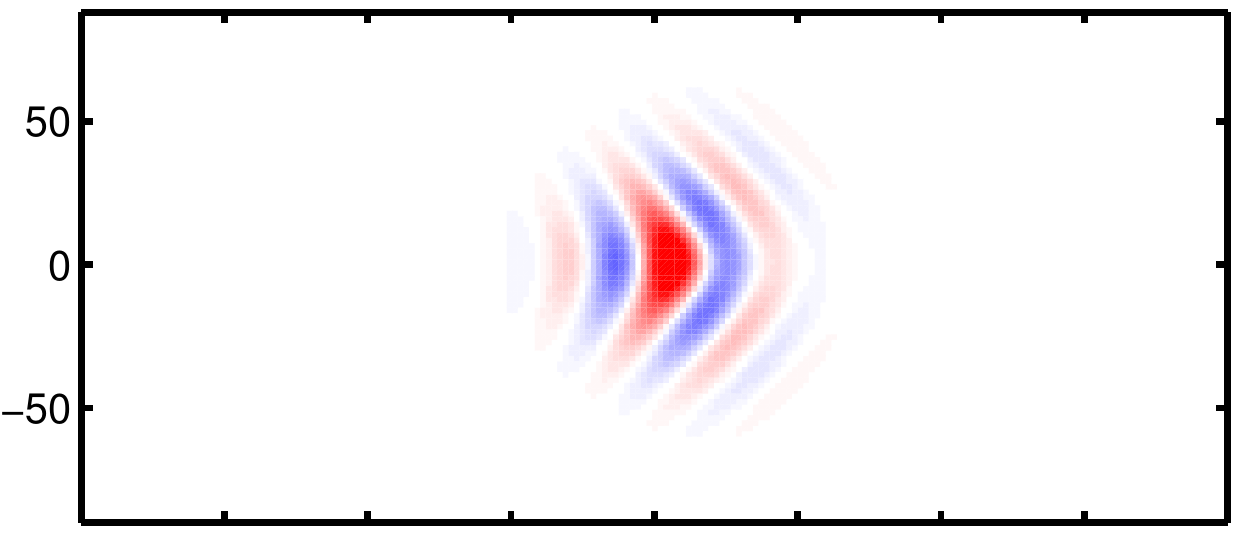}}
\hspace{.25cm}
\subfigure[$t=500$]{\includegraphics[scale=0.45]{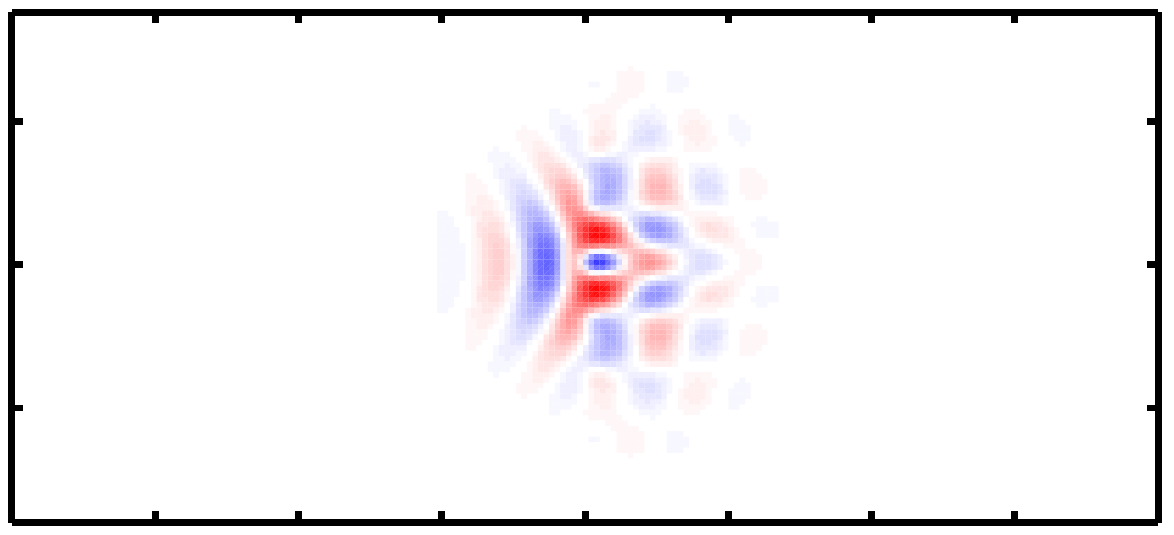}}
\\
\subfigure[$t=800$]{\includegraphics[scale=0.45]{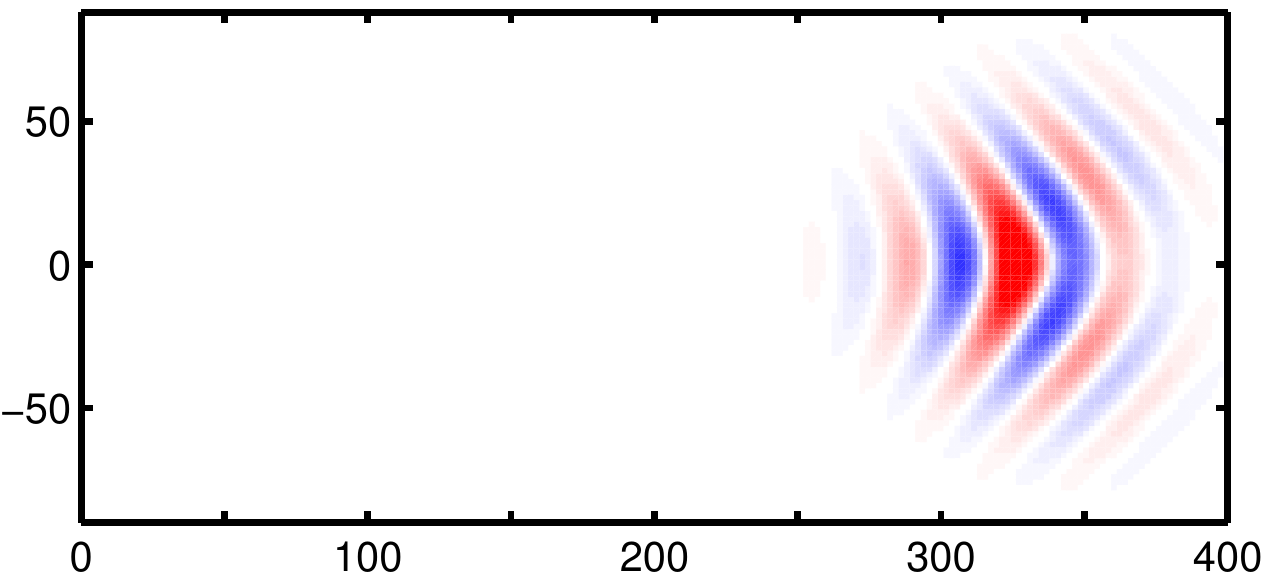}}
\hspace{.05cm}
\subfigure[$t=800$]{\includegraphics[scale=0.45]{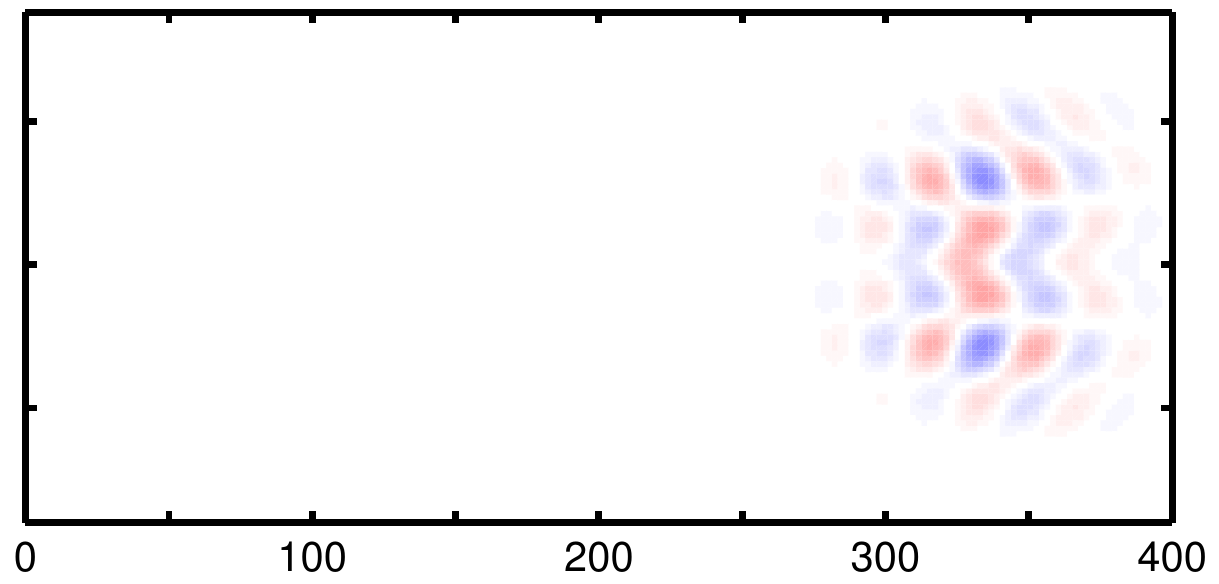}}
\caption{The propagation of a wavepacket in a flow governed by the modified Kuramoto-Sivashinsky equation is shown for a clean configuration (insets $a$-$c$-$e$) and the controlled case (insets $b$-$d$-$f$). The non-dimensional parameters in Eq.~\ref{eq:RP} and Eq.~\ref{eq:RS} are chosen for mimicking the propagation of Tollmien-Schlichting waves at $Re=1000$, with $\alpha_{max} = 1.68\times 10^{-1}$, $\beta_{max} = 2.15\times 10^{-1}$ and $\omega_{max}= 2.67\times 10^{-3}$.}\label{fig:snap}
\end{figure}

In this section, a modified version of the Kuramoto-Sivashinsky equation is used for testing the multi-input case, mimicking the three-dimensional setup sketched in Fig.~\ref{fig1}; this proof-test provides the design feasibility in larger computational domains.

\medbreak
The flow field is approximated in a two-dimensional domain $x$-$z$ plane, and governed by the equation
\begin{equation}
\dfrac{\partial v}{\partial t} = -V\dfrac{\partial}{\partial x}
\left( v -\dfrac{1}{8P}\dfrac{\partial^2 v}{\partial z^2}
\right) 
-\dfrac{1}{R}\left( P\dfrac{\partial^2 v}{\partial x^2} +\dfrac{\partial^4 v}{\partial x^4} +S\dfrac{\partial^4 v}{\partial z^4}
\right),
\end{equation}
With respect to the original equation, the dynamics is linearized around the convective velocity, $V=0.4$. The non-dimensional terms $R$ and $P$ are defined as
\begin{equation}\label{eq:RP}
R = \dfrac{VL^3}{\mu}\qquad\qquad P = \dfrac{\eta L^2}{\mu},
\end{equation}
with $\eta$ representing an energy production term and $\mu$ the dissipation; $L$ is the reference length of the system. The parameters can be chosen ad-hoc such that the dispersion relation of the system mimics the evolution of travelling packets of Tollmien-Schlichting (TS) waves. Introducing the maximum streamwise and spanwise wavenumbers as $\alpha_{max}, \beta_{max} \in \mathbb{R}$ respectively, and the complex temporal frequency $\omega$, the non-dimensional numbers can be expressed as
\begin{equation}\label{eq:RS}
R=\dfrac{P^2}{4\omega_{max}} \qquad\qquad P=\dfrac{2}{\alpha_{max}^2} \qquad\qquad S=\dfrac{\omega_{max} R}{\beta_{max}^4},
\end{equation}
with $S$ providing the modulation along the spanwise direction of the wave. The numbers are chosen such that the dispersion relation closely resembles the behaviour in the horizontal plane of a TS wave evolving on a flat-plate at $Re_{\delta^*}=1000$ (see \cite{semeraro2011feedback}, \cite{semeraro2013transition}). More details are reported by \cite{johan2015controlling}.
\medbreak
The computational box extends along the streamwise direction in the interval $x\in[0, 500]$ and $z\in[-90, 90]$ along the spanwise direction; a grid with $N_x=256$ and $N_z=96$ is chosen. The spatial discretization is performed by means of a pseudo-spectral method, that includes a fringe region extending between $x=400$ and $x=500$; periodic boundary conditions are imposed along the spanwise direction. The time marching is performed using a three-steps Runge-Kutta scheme; the basic implementation can be found in the repository \url{https://github.com/nfabbiane/ks2D}.


\subsection{A multi-input full-order controller} 
\begin{figure}
\subfigure[Control gain, $l=100$]{\includegraphics[scale=0.43]{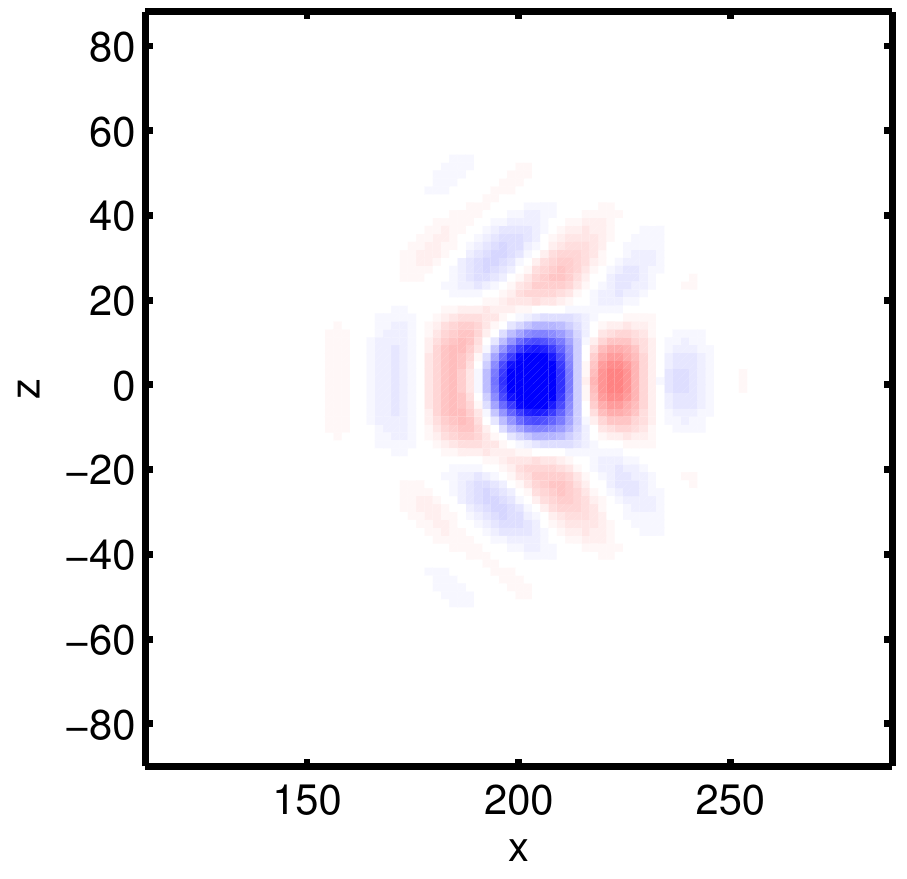}}
\subfigure[RMS of the analysed cases]{\includegraphics[scale=0.41]{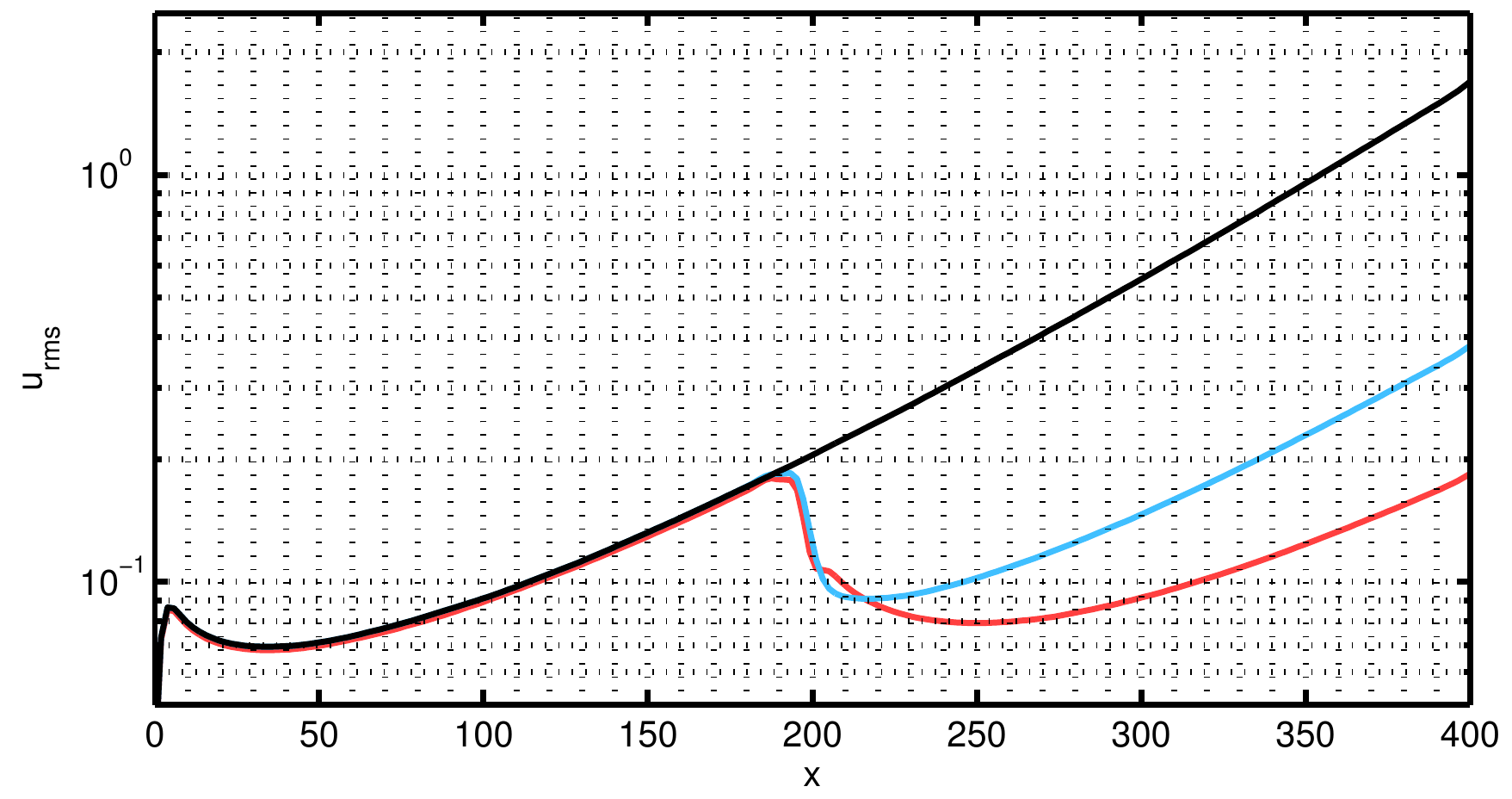}}
\caption{
\emph{Left}: the full-order control gain related to the actuator $\Bb_{\u}$ placed at $\boldsymbol{x}=(200,0)$ using the ADA, with a control penalty $l=100$. \emph{Right}: the root mean squares (RMS) of the velocity is shown as a function of the streamwise direction $x$. The value is computed using $N=10000$ snapshots. The clean configuration (black, solid line) is compared to the controlled cases designed using $l=100$ (red-solid line) and $l=500$ (blue-solid line).
}\label{fig:rms}
\end{figure}
The resulting dynamics is depicted in Fig.~\ref{fig:snap}$a$-$c$-$e$, where the impulse response of the system is shown at three different instants. The disturbance is introduced at $\mathbf{x}_0=\left( x_0, z_0 \right) = \left(2.5, 0\right)$, and its spatial distribution is modelled as
\begin{equation}\label{eq:gau}
\Bb = \exp{\left( -\dfrac{(x-x_0)^2}{\sigma^2} -\dfrac{(z-z_0)^2}{\sigma^2} \right)},
\end{equation}
with $\sigma=4$. The dynamics of the wavepacket mimics the evolution of a TS wave, growing as it propagates downstream along the streamwise direction while extending along the spanwise direction with a backward bending. The corresponding energy growth associated with the evolving wavepacket is shown in Fig.~\ref{fig:rms}$b$ (black, solid line); the root-mean-square energy is obtained from the statistics associated with a stochastically driven simulation, averaged over a span of $T=10000$ time units.

A LQR controller is designed using the ADA algorithm. The controller is centralized: all the gains are designed such that the set of sensors and actuators is coupled. The setup closely resembles the one analysed by \cite{semeraro2011feedback}, \cite{semeraro2013transition} and reproduced schematically in Fig.~\ref{fig1}: a row composed by $n=9$ localized actuators ($\Bb_{\u}$) is placed along the spanwise direction at $x=200$, equi-spaced of $\Delta z =20$. The same setup along the spanwise direction is chosen for the sensors $\Cb_z$, placed at $x=300$. All the chosen elements are modelled as Gaussian distributions, Eq.~\ref{eq:gau}.
Nine iterations are set, one for each control gain. An example is given in Fig.~\ref{fig:rms}$a$, where the control gain placed at $\mathbf{x}=\left(200, 0\right)$, with $l=100$; the controller shows the typical signature of the adjoint solution, as it is bent backward with respect to the propagation of the direct solution. The performances of two controllers are shown in Fig.~\ref{fig:rms}$b$, where the clean configuration is compared with two cases: $l=100$ (red solid line), and $l=500$ (blue solid line). The results are in qualitative agreement with \cite{semeraro2011feedback}. More interestingly, it shows that the technique enables the computation of multi-input, full order controllers without using a preliminary model reduction step.


\section{Conclusions}\label{sec:conclusions}
We extended the framework of the \emph{Adjoint of the adjoint} algorithm to multivariate large systems, by highlighting how decentralized, centralized and coupled controllers can be computed for full-order, optimal control. These observations apply straightforwardly to the dual problem, the estimation problem, and to robust controllers belonging to the $\mathcal{H}_{\infty}$ framework. The main advantage of the ADA algorithm (and its extensions) is the independence of the final solution from the initial conditions. From the physical point of view, this implies that it is not required any knowledge of the disturbances active in the flow. This feature makes the algorithm particularly appealing for the pre-assessment of the optimal performance of a controller, based on linear (or linearized) plant. Moreover, one can assess the controllability of the dynamical system by such techniques and estimate the efficiency of the resulting controller.

\medbreak
Two test cases are considered. A toy-problem mimicking a distributed system is analysed for assessing the convergence performance of ADA against an analogous algorithm by M\aa rtensson \emph{et al.} \cite{maartensson2012scalable}, indicated as MR. The results show a remarkable scalability of ADA, with respect the number of actuators; with respect to the MR algorithm, the number of iterations required is on average three times smaller and - more importantly - it is not affected by the stochastic approach required by the MR algorithm for identifying the global solution. The second example is based on a modified version of the Kuramoto-Sivashinsky. This example reproduces a control setup based on multiple, localized sensors and actuators placed along the cross-flow direction. The feasibility of the approach is demonstrated suggesting that the technique might applied also to full three-dimensional cases in more complex fluid mechanics settings.

\medbreak
Future work will be devoted to the application of these methods in combination with sparsity promoting algorithms (see \cite{2013:lin:fard:jova}). In our application, the coupling among the actuators is pre-determined and not optimized. However, it is possible to introduce the sensor/actuator pairing within the optimization process (pairing problem). Moreover, sensor and actuator placement for large-scale systems might be analysed starting from this application (\cite{chen2011h}, N.~Fabbiane, \emph{private communication}, 2016).

\bigbreak
\noindent \textbf{Acknowledgments} Anders Rantzer and his collaborators (Lund University) are acknoweldged for interesting discussions and for bringing up their work on optimal control synthesis in large systems. Nicol\`{o} Fabbiane (DAFE, Onera) provided the basic implementation of the two-dimensional Kuramoto-Sivashinsky equation. The first author thanks the support by the Agence Nationale de la Recherche (ANR) under the \emph{CoolJazz} project, grant number ANR-$12$-BS$09$-$0024$.

\appendix

%
%
\section{Derivation of ADA algorithm using integration by parts}\label{app:A}
In this section we briefly summarize the original derivation of the ADA algorithm, as first proposed by \cite{2016:bewley:luchini:pralits}. We start from the direct-adjoint system
\begin{subeqnarray}\label{eq:ada_luc1}
\dfrac{d \qv}{dt} &=& \Ab\qv -\Bb\Rb^{-1}\Bb\pv, \quad \qquad with \quad \qv(0)=\qv_0, \\
\dfrac{d \pv}{dt} &=&-\Ab^H\pv -\Qb\qv, \,\,\, \qquad \qquad with \quad \pv(T)=0, 
\end{subeqnarray}
where the optimality condition has already been included. By introducing two variables, namely the state vectors $\av(t)$ and $\bv(t)$, we can recombine the system as
\begin{eqnarray*}
\int_0^T\boldsymbol{a}^H\left(
\dfrac{d \qv}{d t}
-\Ab\qv +\Bb\Rb^{-1}\Bb^H\pv\right)dt +
\int_0^T\bv^H
\left(
\dfrac{d \pv}{d t}
+\Ab^H\pv +\Qb\qv\right)dt.
\end{eqnarray*}
Integration by parts allows rewriting the relation as
\begin{eqnarray*}
\int_0^T\qv^H\left(
-\dfrac{d \av}{d t}
-\Ab^H\av +\Qb\bv\right)dt +
\int_0^T\pv^H
\left(
-\dfrac{d \bv}{d t}
+\Ab\bv +\Bb\Rb^{-1}\Bb^H\av\right)dt + \\
(\av^H\qv+\bv^H\pv)_T -
(\av^H\qv+\bv^H\pv)_0
\end{eqnarray*}
Gathering the resulting state equations for $\av(t)$ and $\bv(t)$, and changing the variables $\av \rightarrow \tilde{\pv}$, $\bv \rightarrow -\tilde{\qv}$, allow us to write the following system
\begin{subeqnarray}\label{eq:ada_luc2}
\dfrac{d \tilde{\qv}}{d t} &=& \Ab \tilde{\qv} +\Bb\Rb^{-1}\Bb^H\tilde{\pv}, 
\qquad\,\,\, \tilde{\qv}(0)=\tilde{\qv}_0,
\\
-\dfrac{d \tilde{\pv}}{d t} &=& \Ab^H\tilde{\pv} +\Qb{\tilde{\qv}}, 
\qquad\qquad\quad\, \tilde{\pv}(T)=0,
\\ 
\tilde{\pv}(T)^H\qv(T) 
-\tilde{\qv}(T)^H\pv(T)
&=&
\tilde{\pv}(0)^H\qv(0) 
-\tilde{\qv}(0)^H\pv(0).
\end{subeqnarray}
Equations \ref{eq:ada_luc2} correspond to the ones of the original optimal control problem \ref{eq:ada_luc1}. The last relation \ref{eq:ada_luc2}$(c)$ is exactly the condition recovered by symplectic product in Eq.~(\ref{eq:symplet}); thus, the observations done in Eqs.~(\ref{eq:symplet}--\ref{eq:adaopt}) are valid also for this case.

%
\section{Full-order controllers using stochastic gradients}\label{app:B}
In this appendix, the algorithm discussed by M{\r a}rtensson and Rantzer in \cite{maartensson2012scalable} is derived following the Lagrangian approach. For sake of conciseness, we will indicate the algorithm with the acronym MR. The state equation is re-written as
\begin{eqnarray}
\dfrac{d \qv}{dt} &=& (\Ab +\Bb\Kb)\qv \quad \qquad with \quad \qv(0) =\qv_0,
\end{eqnarray}
The problem is rewritten now assuming the control kernel $\Kb$ as unknown of the problem, replacing $\u(t)$. The cost function is now
\begin{eqnarray}
\mathcal{J} = \dfrac{1}{2}\int^T_{0}\left(\qv^H \Qb \qv +\qv^H \Kb^H \Rb \Kb \qv\right)dt =
\dfrac{1}{2}\int^T_{0} \qv^H \left(\Qb +\Kb^H \Rb\Kb\right)\qv\,dt,
\end{eqnarray}\label{eq:lyap}
and the resulting augmented Lagrangian is
\begin{eqnarray*}
\mathcal{\tilde J} =\mathcal{J}-\int^T_{0}\pv^H\left(\dfrac{d \qv}{dt} -\Ab\qv -\Bb\Kb\qv \right)dt.
\end{eqnarray*}
The resulting system is obtained by considering the gradients with respect to $\pv$, $\qv$ and $\Kb$
\begin{subeqnarray}\label{eq:sysK}
\dfrac{\partial \mathcal{\tilde{J}}}{\partial\pv}&=& \dfrac{d \qv}{dt} -\Ab\qv -\Bb\Kb\qv, \\
\dfrac{\partial \mathcal{\tilde{J}}}{\partial\qv}&=& \dfrac{d \pv}{dt} +\Ab^H\pv +\Kb^H\Bb^H\pv
+\left(\Qb +\Kb^H \Rb \Kb \right)\qv, \\
\dfrac{\partial \mathcal{\tilde{J}}}{\partial{\Kb}}&=& \int_0^T\left(\Rb\Kb\qv +\Bb^H\pv\right)\qv^H\,dt.
\end{subeqnarray}
Zeroing the gradients, the system can be arranged in matrix form as
\begin{displaymath}
\dot{ \left(\begin{array}{c}
\qv \\
\pv \end{array}\right)}
= \left[\begin{array}{cc}
\Ab +\Bb\Kb & \boldsymbol{0} \\
-\left(\Qb+\Kb^H\boldsymbol{R}\Kb\right) & -\left(\Ab +\Bb\Kb\right)^H \end{array}\right]
\left(\begin{array}{c}
\qv \\
\pv \end{array}\right).
\end{displaymath}\label{eq:MRmat}
The solution is updated by using $\ref{eq:sysK}c$ as 
\begin{equation}
\Kb^{i+1}=\Kb^{i} -\beta\nabla_{\boldsymbol{K}}\mathcal{\tilde{J}}.
\end{equation}
The iteration is usually initialized with a guess for $\boldsymbol{K}$, unless the system under consideration is asymptotically stable; in the latter case, the controller is designed for modifying the transient dynamics and the starting guess is simply a null vector. In this formulation, the problem depends on the initial conditions. The solution is not guaranteed to be the one corresponding to the global minimum for a given initial condition. This drawback can be circumvented by using \emph{ad-hoc} strategies as the stochastic gradient approach described in \cite{maartensson2012scalable}. Moreover, with respect to the standard algorithm, the MR algorithm does not require the solution of Lyapunov equations, replaced with an approximation based on the adjoint and direct solutions.

%
\subsection{Relation with the ADA algorithm: generalizing the sensitivity approach}
The symplectic product adopted in ADA can be introduced also for the system in Eq.~\ref{eq:MRmat}. It is possible to show that the relation
\begin{eqnarray}\label{adaMR}
\tilde{\pv}(t)^H{\qv}(t) = \tilde{\qv}(t)^H{\pv}(t) \qquad \forall t,
\end{eqnarray}
is fulfilled also in this case.
Considering again the gradient, 
\begin{eqnarray}
\nabla_{\Kb} \mathcal{\tilde{J}} &=& \int_0^T\left(\Rb\Kb\qv +\Bb^H\pv\right)\qv^H\,dt,
\end{eqnarray}
it is null $\forall \qv$ if
\begin{eqnarray}
\Rb\Kb\qv(t) = -\Bb^H\pv(t).
\end{eqnarray}
By rescaling the relation via $\Rb^{-1}$, we recover the equivalence between the product and the optimality condition at $t=0$
\begin{subeqnarray}
\tilde{\pv}(0)^H{\qv}_0 &=& \tilde{\qv}(0)^H{\pv}_0, \\
\Kb\qv_0 &=& -\Rb^{-1}\Bb^H\pv_0.
\end{subeqnarray}
Thus, also for the basic equation of the MR algorithm, introducing as initial conditions of the system $\boldsymbol{\tilde{q}}=-\boldsymbol{R}_{i,i}^{-1}\Bb_i$, the iteration enables to compute the $i$-th kernel $\boldsymbol{K}$ as solution of the adjoint equation at the final time $T$. All the observations done for the multi-input, coupled version of the ADA algorithm are valid also for this algorithm. Interestingly, this algorithm produces the control kernels in two different ways: i) as adjoint solution; ii) as result of the gradient-based iteration
\begin{equation}
\boldsymbol{K}^{n+1}=\boldsymbol{K}^{n} -\beta\nabla_{\boldsymbol{K}}\mathcal{J}.
\end{equation}
The sensitivity with respect to the initial conditions applied to the optimal control problem leads to the same results with different choice of gradients. The gradient with respect to $\boldsymbol{K}$ can be rewritten, using the chain-rule, as 
\begin{eqnarray}
\dfrac{\partial \mathcal{\tilde{J}}}{\partial \Kb}=\dfrac{\partial \mathcal{\tilde{J}}}{\partial{\u}}
\dfrac{\partial \mathcal{{\u}}}{\partial{\Kb}}=\left(\Rb\Kb\qv +\Bb^H\pv\right)\qv^H.
\end{eqnarray}
The underlying system of equations is equivalent in the two cases.


\bibliographystyle{elsarticle-num}
\bibliography{refs}

\end{document}